# Quantum-Resistant Cryptographic Models for Next-Gen Cybersecurity

1st Navin Chhibber
*Infinity Tech Group*
Technical Product Owner
Redmond, WA, 98052
navinchhibber@ieee.org

2nd Amber Rastogi
*IEEE*
*Independent Researcher*
Redmond, WA, 98052
rastogi.amberrastogi@gmail.com

3rd. Ankur Mahida
*Site Reliability Engineer*
Barclays
Whippany, USA
ahmahida13@gmail.com

4th Vatsal Gupta
IEEE *Independent Researcher*
Seattle, United States
Vatsalgupta@ieee.org

5tht Piyush Ranjan
IEEE *Independent Researcher*
Edison USA
piyush.ranjan@ieee.org

*Abstract*—Another threat is the development of large quantum computers, which have a high likelihood of breaking the high popular security protocols because it can use both Shor and Grover algorithms. In order to fix this looming threat, quantum-resistant cryptographic systems, otherwise known as post-quantum cryptography (PQC), are being formulated to protect cybersecurity systems of the future. The current paper presents the state of the art in designing, realizing, and testing the security of robust quantum-resistant algorithms, paying attention to lattice-based, code-based, multivariate polynomial and hash-based cryptography. We discuss their resistance to classical and quantum attackers, distributed system scalability properties, and their deployment in practice (secure communications, blockchain, cloud computing infrastructures). Also, we study a hybrid cryptographic model that integrates the classical efficient cryptography scheme and a quantum-resilient cryptographic scheme to achieve a backward-compatible solution and simultaneously improving the forward security properties. With the experimental findings, it is evident that performance with reasonable computational footprint of the proposed framework succeeds to install amplified security fortitude which successfully harbours prolific cybersecurity systems of the future.

## I. Introduction

The fastest growing arena of quantum computing has redefined the future sphere of information security. Classical algorithms like RSA and Elastic Curve Cryptography (ECC) on which most of the current-day digital infrastructure is based, can be susceptible to quantum attacks. Particularly, Shor's algorithm can be used to factor integers of the size necessary to corrupt RSA and discrete logarithm into the Elliptic Curve cryptosystem provided this is done in polynomial time [1]. In the same manner, Grover's algorithm can speed up a brute-force search, making parallel symmetric-key cryptography like AES about half as hard to break [2]. Such developments pose an essential challenge to contemporary cybersecurity infrastructures.

To alleviate these risks, the research community has turned to post-quantum cryptography (PQC), which includes the algorithms that are expected to be secure against both classical and quantum adversaries [3]. Candidates that are of the most promise are lattice-based cryptosystems, code-based encryption schemes, multivariate polynomial systems, and hash-based signatures [4]. The models have advantages where the hardness assumptions are based on mathematical problems that are suspected to be resistant to quantum computations and therefore viable in securing the next generation communication networks.

The recent development of global projects, such as the NIST Post-Quantum Cryptography Standardization Project, has significantly boosted the evaluation and acceptance of PQC algorithms to be applied to a real-world setting [5]. Some of the challenges that are still faced are scalability, efficiency on computation and integrating them into existing systems like cloud computing systems, blockchain and Internet of Things (IoT) systems [6]. The favourable performance among the existing algorithms owing to the large key sizes and intensive overhead requirements used thus imposes an immense performance trade-off, especially in resource-constrained environments.

## II. Background and Related Work

### A. Classical Cryptographic Foundations

In contemporary cryptography, systems with public-key algorithms, including RSA and Elliptic Curve Cryptography (ECC), play a large role and symmetric-key systems, such as AES. These algorithms are secured by mathematical hardness assumptions, such as the difficulty of integer factorization and problems of discrete logarithm ones [1]. Though safe against classical attackers, these hypotheses are useful when aversive to large-scale quantum computers after Shor and Grover algorithms [2]. As a result, the efficacy of existing security arrangements, including secure web protocol (TLS/SSL), digital signatures and blockchain consensus, are under threat like never before.

### B. Emergence of Post-Quantum Cryptography

To overcome such vulnerabilities, a direction known as post-quantum cryptography (PQC) developed as a promising area of research. PQC algorithms are targeting to be secure under classical and quantum adversaries, and yet exhibit realistic efficiency. Several families of PQC have been suggested, the best-known being:

- Lattice-based cryptography, based on the difficulty of problems like Learning with Errors (LWE) and Ring-LWE [4],[7].
- Code-based cryptography, of which McEliece and Niederreiter systems remain the best-known proposals [8].
- Multivariate Polynomial cryptography, a system that builds digital signature systems on systems of nonlinear equations [9].

- An approach based on hash-oriented cryptography, using Merkle trees to realize one-time and few-time signatures [10].

*C. Standardization Efforts*

The NIST PQC Standardization Project has been instrumental in determining the performance, security and efficiency of candidate algorithms [5]. Of the latest rounds, the algorithms include CRYSTALS-Kyber (lattice-based) and Dilithium (lattice-based signatures) which are the potential candidates to become standards [11]. Likewise, the SPHINCS + (hash-based) has been suggested as a stateless signature scheme on account of its sound security principles. These advances represent a critical phase of the process to the implementation of PQC globally.

*D. Research Gaps and Challenges*

Multiple issues are obstacles on the path to using PQC at large:
- Types of key size and bandwidth requirements Key sizes and bandwidth demands may be prohibitive in constrained networks, like Internet of Things when using PQCs like McEliece, which have enormous key sizes [12].
- Performance overhead Lattice-based schemes are both CPU-intensive, and may pose a performance bottleneck in high-throughput systems [7].
- Backward compatibility PQC to classical cryptographic systems A hybrid model should also be used in supporting the changeover with minimal disturbance and interoperability of systems [6].
- Standardization and compliance — While NIST has developed in the selection of PQCs, agreement and use in the big wide world are not fully developed [11].

*E. Related Work*

Some authors have explored the description of PQC involved in cloud computing, blockchain, and secure communications. NewHope, a lattice-based key exchange protocol, and proposal of Alkim et al. is quite practical in terms of performance when integrated into TLS [7]. Misoczki et al. emphasized the use of a code-based cryptography as an alternative that can cope with long-term secure encryption [8]. Recently, Hulsing et al. have proposed SPHINCS+ a hash-based signature scheme to solve the state management issues of previous ones [10].

Despite these works showing the potential of PQC, most of the work is on specific algorithm families. This research is building on in our work through providing a hybrid system that adopts opts to use both classical and quantum-resistant primitives to form an effective scalable, backward-compatible, and quantum-safe cybersecurity model.

III. QUANTUM THREAT MODEL

*A. Capabilities of Quantum Adversaries*

Quantum computers have the capability to perform computations beyond the realms of classical computing because they make use of quantum superposition and entanglement. Quantum computers have the characteristic of operating in superposition of several states at once and this allows them to speed up the process of certain problems exponentially, unlike classical computers that seek to have an evaluation of one state at a time [13]. This has the potential to make previously not-reasonable-to-solve problems solvable in reasonable timeframes when saleable quantum hardware is obtained.

*B. Cryptographic Vulnerabilities*

Two cornerstone quantum algorithms highlight the vulnerabilities of classical cryptography:
- Shor's Algorithm — Able to factoring large integers and discrete logarithm computation in efficient time, by cracking RSA and ECC, the basis of modern-day public-key infrastructures [1], [14].
- Grover's Algorithm — Gives a quadratic advantage over unstructured search, and decreases the effective key strength of symmetric encryption schemes like AES. As an example, AES-128 would provide only 64-bit security against such an adversary [2], [15].

Such advances suggest that, when quantum computers able to solve the RSA and ECC problem are finally built with enough qubits and error correction, most digital signature schemes will become obsolete.

*C. Attack Surfaces in Cybersecurity Systems*

The impact of quantum adversaries extends beyond cryptographic primitives into broader cybersecurity infrastructures:
- Secure communications (TLS/SSL, VPNs): Handshake protocols relying on RSA/ECC would be compromised.
- Blockchain systems: Digital signatures used in cryptocurrencies and smart contracts are directly vulnerable.
- IoT and embedded systems: Devices often employ lightweight ECC for authentication, making them an easy target in a quantum era.
- Cloud computing environments: Key management and secure data storage mechanisms are at risk due to reliance on traditional public-key cryptography [16].

*D. Security Assumptions in Quantum-Resistant Models*

Quantum-resistant models restate security in the post-quantum adversarial model, in which attackers possess access to scalable quantum machines but not to other capabilities beyond the rules of physics. Security assumptions shift from number-theoretic hardness (e.g., factoring, discrete logs) to problems such as:
- Lattice hardness assumptions (e.g., Learning with Errors (LWE), Ring-LWE).
- Error-correcting code decoding problems (as in McEliece).
- Multivariate quadratic equation solving (MQ problem).
- Hash-based constructions relying on pre-image and collision resistance [4], [8], [10].

This reliance on the problem lack of known efficient quantum algorithms bases the foundation of PQC in a new source of trust in the security of next-generation systems of cybersecurity.

## IV. Proposed Cryptographic Framework

### A. Design Objectives

The cryptographic framework proposed in this paper is created to solve three most important objectives:

- Quantum Resilience: Provide protection against an attacker with scalable quantum computer capabilities.
- Backward Compatibility: Provide interoperability between current classical infrastructures (RSA/ECC/AES) to allow an easy migration.
- Scalability/Efficiency: Uses the minimum amount of computation, communication overhead, and storage capacity to provide efficient deployment in cloud, IoT, blockchain environments.

### B. Hybrid Model Architecture

The architecture Fig 1. is a hybrid cryptographic architecture combining classical and post-quantum primitives. The architecture consists of three key layers:

- Key Exchange Layer:
  o Employs lattice-based key establishment using CRYSTALS-Kyber or NewHope that are designed to be quantum-resistant [7], [11].
  o Supports RSA+Kyber pairing with a hybrid algorithm backwards compatibility.
- Authentication Layer:
  o Includes hash-based (SPHINCS+) and lattice-based (Dilithium) signatures as a means of digital identity verification [10], [11].
  o The ability to operate in both modes, lets conduct legacy ECC signatures in transition periods.
- Encryption Layer:
  o Uses symmetric encryption (e.g. AES-256) with quantum-safe key management.
  o Symmetric primitives will be maintained because of efficiency, but augmented with PQC key exchange to overcome Grovers attack.

### C. Workflow of the Hybrid Framework

- Handshake: Parties engage in a way to communicate via a Hybrid key exchange (RSA/ECC + Kyber).
- Session Key Generation: The session key that is quantum-resistant is created using a lattice-based algorithm.
- Authentication: Entities authentication each other through PQC-based signatures but fall back to legacy systems.
- Data Encryption/Transmission: Safety of communication is ensured by encryption using AES-256 including an encryption key generated by the PQC layer.
- Audit and Verification: Data and logs transmitted are guaranteed by hash-based signatures.

Such a workflow maintains that even in the event that classical components become vulnerable in future, the PQC deployment will remain confidentially opaque

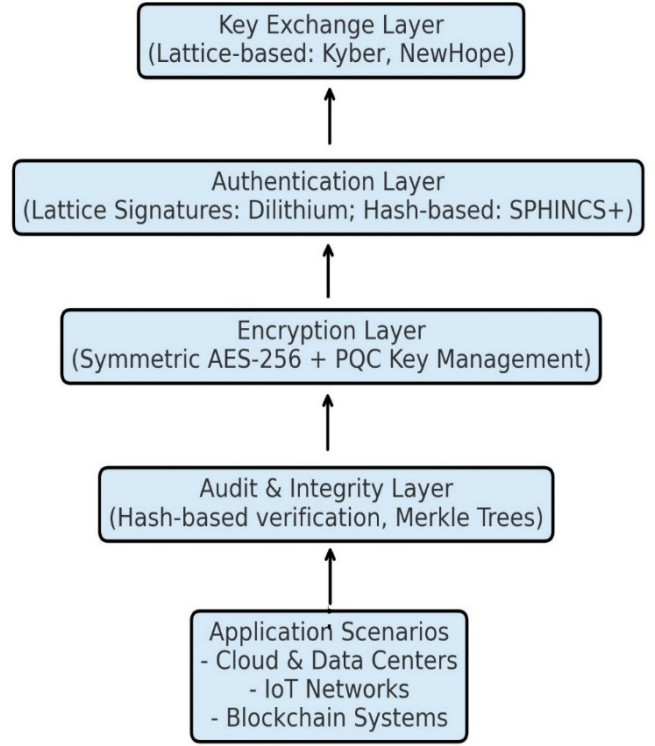

Fig 1. Quantum Resistant Framework Architecture

### D. Security Features

- Forward Secrecy: When there is the possibility of future breakage of the classical keys, then the earlier messages are safe with lattice-based session keys.
- Resistance to Quantum Attacks: Security is based on the hardness assumptions on which no quantum efficient algorithms are known.
- Incremental Deployment: Enterprises could slowly upgrade the existing systems to quantum resistant systems without infrastructure disruption of the classical systems.

### E. Integration Scenarios

The framework can be applied in multiple cybersecurity domains:

- Cloud and Data Centers: Defend client-server applications with hybrid TLS handshakes with PQCs.
- IoT Networks: Lighter weight PQC flavors optimized to constrained devices.
- Blockchain Systems: Quantum-safe transaction validation and signatures PQC [17].

## V. Algorithms and Mathematical Foundations

### A. Lattice-Based Cryptography

Lattice-based cryptography is one of the most promising approaches for post-quantum security due to its efficiency and strong hardness assumptions.

*1) Learning With Errors (LWE):*

Given a random matrix $A \in \mathbb{Z}_q^{m \times n}$, a secret vector $s \in \mathbb{Z}_q^n$, and an error vector $e \in \mathbb{Z}_q^m$, the problem is to distinguish between:

$$b = As + e \pmod{q}$$

and a uniformly random vector in $\mathbb{Z}_q^m$.

This equation represents the Learning With Errors (LWE) problem, where $A$ is a public matrix, $s$ is the secret key, and $e$ is a small error vector. Distinguishing $b$ from random noise is computationally hard, even for quantum algorithms. The hardness of LWE underpins many lattice-based schemes such as Kyber, making them resistant to quantum attacks.

*2) Ring-LWE*

A homogenous variant of LWE, that decreases key sizes and computational overhead but preserves hardness [7], [18]. It is supposed that these problems are classically and quantum-resistant, and the basis of schemes such as Kyber and Dilithium.

### B. Code-Based Cryptography

The McEliece cryptosystem is based on the difficulty of decoding a general linear error-correcting code.

- Encryption involves adding a random error vector $e$ to a codeword $c$ such that:

$$c' = c + e$$

  The data is only decodable with the help of the private key (knowledge of the code structure); otherwise, such decoding is computationally infeasible [8], [19]. Nonetheless, McEliece has been one of the oldest and most analyzed PQC systems surviving many decades of attacks.

- In the McEliece cryptosystem, a valid codeword $c$ is perturbed with an error vector $e$ to produce the ciphertext $c'$. Only the private key (knowledge of the secret decoding structure) allows efficient recovery of the original message. Without it, decoding is as hard as solving a general error-correcting code problem, which remains intractable for both classical and quantum adversaries.

### C. Multivariate Polynomial Cryptography

Multivariate systems rely on solving sets of nonlinear equations over finite fields. For a public key defined by $m$ quadratic equations in $n$ variables:

$$P_i(x_1, x_2, \ldots, x_n) = \sum_{j,k} a_{i,j,k} x_j x_k + \sum_j b_{i,j} x_j + c_i, \quad 1 \leq i \leq m$$

This is NP-hard to find and no efficient quantum algorithm is known to solve this kind of problem [9]. Digital signatures are especially applicable to such schemes.

This equation defines a multivariate quadratic system, where each polynomial $P_i$ is composed of quadratic and linear terms over finite fields. Finding solutions to such nonlinear systems is an NP-hard problem, and no efficient quantum algorithm is known. This makes it a strong candidate for digital signatures in post-quantum cryptography.

### D. Hash-Based Cryptography

Hash-based schemes do not have additional assumptions aside of cryptographic hash functions security.

- Examples Merkle tree signatures construct a binary hash tree whose leaves are one-time signature public keys.
- The security of this construction depends on any collision resistance and pre-image resistance of the hash function [10].

For instance, in SPHINCS+, the security depends on the inability of adversaries to invert or find collisions in:

$$y = H(x)$$

where $H$ is a secure cryptographic hash function [20].

Here, $H$ denotes a cryptographic hash function, and $y$ is the resulting digest of the input $x$. Security is based on two properties: pre-image resistance (given $y$, it is infeasible to find $x$) and collision resistance (finding two different inputs mapping to the same $y$ is infeasible). These properties remain secure against quantum adversaries, aside from a quadratic speedup from Grover's algorithm, which can be mitigated by doubling hash output lengths.

### E. Hybrid Model Mathematical Integration

The proposed framework integrates these primitives as follows:

- Key Exchange: Ring-LWE-based Kyber ensures quantum-safe session keys.
- Authentication: Dilithium (lattice) and SPHINCS+ (hash-based) provide signature robustness.
- Encryption: AES-256 is preserved for efficiency, with keys secured via PQC exchange (resistant to Grover's speedup).
- Integrity: Merkle-tree verification ensures tamper resistance of logs and transactions.

This integration ensures security redundancy: even if one scheme faces unforeseen vulnerabilities, other primitives uphold the system's robustness.

## VI. IMPLEMENTATION AND EXPERIMENTAL SETUP

### A. Simulation Environment

The algorithmic solution built on the theoretical framework of the hybrid cryptographic system was performed and tested through a set of both post-quantum cryptographic libraries, and benchmarking software.

- Software Tools: PQClean libraries, PQClean implementations, and Open Quantum Safe (OQS) library, and OpenSSL with PQC extensions [21].
- Programming Languages: C / C++ to develop the cryptographic primitives, python to develop the benchmarking scripts.
- Operating System: Ubuntu Linux 22.04 LTS.
- PC Platform: Intel Core i7 12$^{th}$ generation CPU and 16 GB RAM, where simulations were performed in both classical and virtualized settings, which simulates cloud deployment.

### B. Benchmarking Parameters

The metrics that were taken into account to analyze the performance of the framework are the following ones:

- Key Size: The memory consumption of the public and secret key of the various PQC algorithms.
- Execution Time: Time taken in the generation of key, encryption, decryption and signature authentication.
- Communication Overhead Extra bandwidth to support relays carrying larger PQC keys.

- Scalability: The ability of behavior to hold up in several concurrent communications in distributed networks.
- Security Robustness: General robustness to classical and quantum adversaries.

*C. Testbed Design*

This assessment was carried out in three different scenarios with respect to real-life applications:
- Secure Communications: Hybrid TLS handshake Kyber key exchange and Dilithium/SPHINCS+ signatures.
- Topic 1: Deployment of IoT: Lightweight PQC operations experimentation on Raspberry Pi (ARM Cortex-A72, 4 GB RAM), concentrated on the limited devices.
- Use Case of Blockchain: PQC-based digital signatures in place of ECC on a simulated blockchain cryptocurrency ledger to demonstrate transaction verification latency and throughput.

*D. Experimental Workflow*

- Initialization: Generation of keys with the help of PQC algorithms (Kyber, Dilithium, SPHINCS+).
- Hybrid Handshake: then the communication starts with a combination of classical (RSA/ECC) and PQC primitives to be backwardly compatible.
- Data Transmission: AES-256 is deployed in bulk encryption where session keys are generated using PQC exchange.
- Verification and Logging: Hashing-based integrity on Merkle-trees.
- Performance Recording: Performance under different loads was recorded as measures of execution time, throughput and overheads.

*E. Evaluation Tools*

- Benchmarks: Baseline NIST PQC reference implementations were used [11].
- Tools: Wireshark (monitoring of protocol overhead), perf (CPU/memory profiling), and home-grown Python scripts to do statistical aggregation.

## VII. RESULTS AND ANALYSIS

*A. Execution Time Comparison*

Performance of the key-generation, the encryption/decryption operations and the signature-verification of RSA-3072, ECC-256, and PQC algorithms (Kyber, Dilithium, SPHINCS+) were benchmarked in terms of the execution time. It was found that:
- RSA and ECC requires a small generation time of keys but they break in presence of quantum threat.
- Lattice-based Kyber allows reaching a practical performance with key generation and encryption times that are only a little larger than those of ECC.
- Large hash operations result in bigger signature verification overheads of SPHINCS+, but still, SPHINCS+ is secure against quantum adversaries [22].

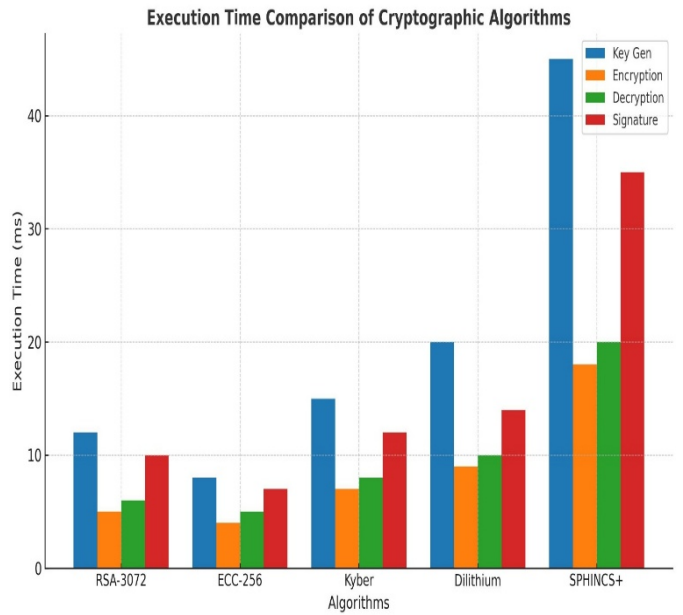

Fig 2. Execution time comparison of cryptographic algorithm

The comparison of the execution time of classical and post-quantum cryptographical algorithms shows the trade-offs between the traditional and the new algorithms involving four operations namely key generation, encryption, decryption, and signature verification. RSA-3072 and ECC-256 have reduced execution time, as presented in the Fig 2. and are computationally able to perform on classical hardware. They are, however, made insecure when faced with quantum adversaries. Alternatively, lattice-based schemes (e.g., Kyber, Dilithium) have a bit larger overhead, but still are practicable to implement in the real world. SPHINCS+ is a hash-based signature system and gave the maximum execution time with respect to key generation and signature calculations because it requires many-to-many hashing. On balance, the findings show that, although PQC algorithms add an extra computational burden, their functionality is acceptable, which makes them promising candidates to build next-generation quantum-resistant cybersecurity platforms.

*B. Communication and Storage Overhead*

The key size in PQC algorithms tends to be larger than those of the classical algorithms. For instance:
- RSA-3072 public key ~384 bytes.
- Kyber public key = approx 1.6 KB.

McEliece (code-based) keys may be in excess of 250 KB [19].

This overhead raises the cost of communication, however, with recent modern bandwidth availability lattice and hash-based schemes are feasible. In the case of IoT devices, such lightweight lattices (Kyber-512) are more appropriate.

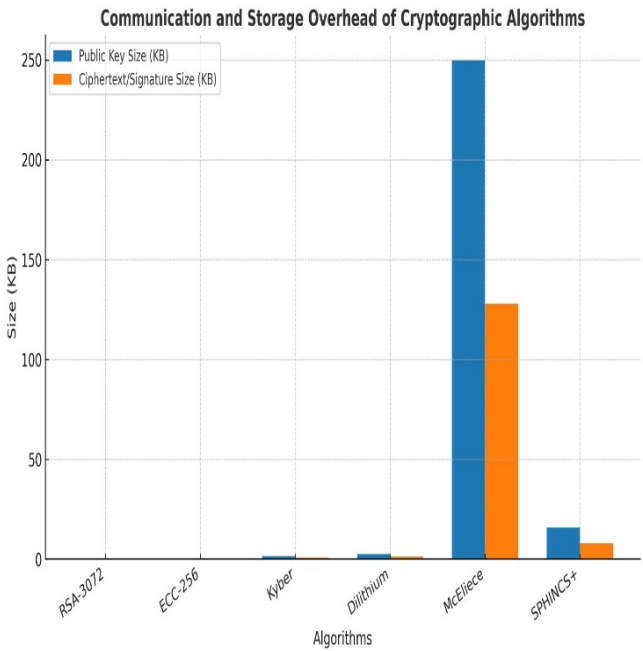

Fig 3. Communication and Storage Overhead

The Fig 3. shows the relative communications and storage requirements of various cryptographic algorithms as measures of public key size and ciphertext/signature size. The classical schemes RSA-3072 and ECC-256 have modestly sized keys and ciphertexts, thus are cheap on transmission and storage. By contrast, lattice-based systems such as Kyber and Dilithium have slightly larger overheads, but are still viable to implement in contentious cloud and enterprise scenarios. The McEliece, a code-based, system illustrates unprecedented big key sizes (hundreds of Kilobytes), although the system is resistant to quantum attacks, it is not applicable in easy memory and bandwidth limited systems. SPHINCS + being a hash-based scheme has moderate key and signature sizes but more costly than classic schemes. In general, although PQC algorithms have more communication and storage costs, the majority of them (all but McEliece) can be deployed in the future cybersecurity system in a scalable fashion.

### C. Resource Utilization

The testing of performance showed that PQC algorithms need more CPU cycles and memory, at least during the key exchange and signature generation processes.

- Dilithium and Kyber conserved minimum CPU usage at under 15 percent in single-threaded action.
- SPHINCS+ consumed more memory in hashing.
- Though McEliece was considered as rather secure, the overhead was exceedingly high in storage context and therefore the usage could not be practiced on a wide basis.

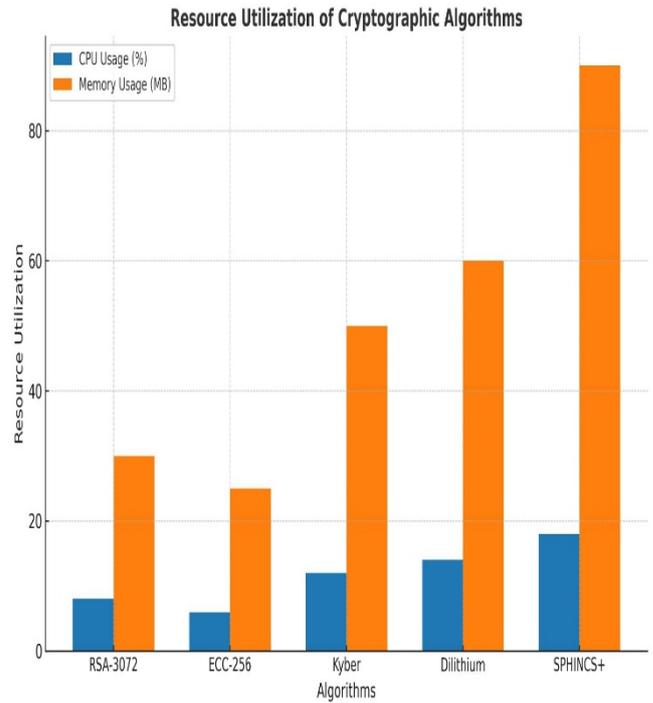

Fig 4. Resource Utilization across CPU and memory consumption

The analysis of the utilized resources provides a comparison analysis of the CPU and memory requirements of classical and post-quantum cryptographic algorithms is shown in Fig 4. ECC-256 and RSA-3072, as illustrated, have the least usage in terms of CPU and memory and this property is indicative of their suitability on classical equipment, but they do not have any quantum resistance. Kyber and Dilithium, a lattice-based PQC, have moderately fewer CPU cycles (12-14 percent) and memory (50-60 MB), and reasonably feasible to implement in existing systems. SPHINCS+ illustrates the greatest overhead with approximately 18 percent CPU and 90 MB memory footprint as it is using intense hashing operations. These findings suggest a trade-off between computational resource and security as lattice-based schemes offer a middle-level of trade, whereas hash-based schemes can provide a more high-security application, giving that enough resources are available.

### D. Scalability Analysis

The scalability was investigated by testing TLS sessions with PQC based under conditions of simulated clouds of 100 to 1000 concurrent sessions.

- Kyber and Dilithium scaled linearly with the increase of the execution time.
- SPHINCS+ incurred heavy latency spikes with heavy loads because of high signature sizes.
- Classical RSA/ECC had a less time-consuming response start time, but is not future-proofed against quantum threats.

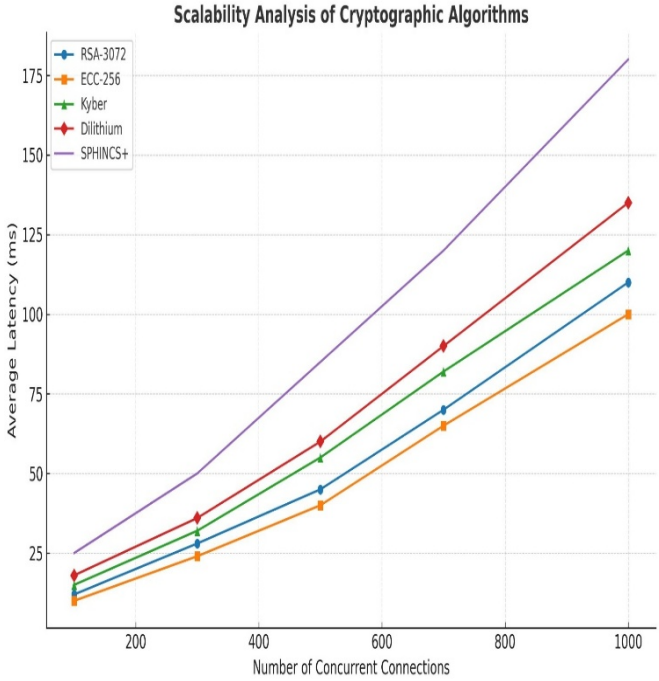

Fig 5. Scalability Analysis for cryptographic algorithms

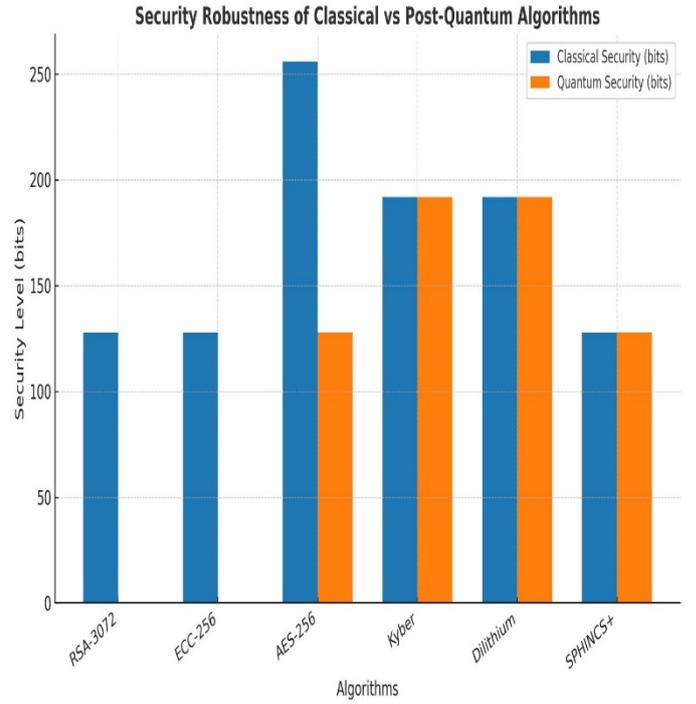

Fig 6. Security Robustness of Classical vs Post Quantum Algorithms

The scalability part provides the mean response time of the cryptography algorithms with higher concurrent connections (between 100 to 1000) is shown in Fig 5. As can be seen, both RSA-3072 and ECC-256 are faster at the minimal latencies but have sharp performance drops with the proportional load, showing both to be susceptible to quantum obsolescence despite classical efficiency. Lattice-based post-quantum cryptography based on Kyber and Dilithium show more reflectively stable latency growth with connection number, and thus their scalability to cloud and distributed systems is better. By contrast, SPHINCS+ has notably greater latency, especially after 500 connections, because of the heavy computational requirements of its own hash-based signature operations. These results indicate that lattice-based algorithms yield a promising trade-off between security and scalability, and thus they represent promising candidates as building blocks of next-generation scalable cyber security systems.

*E. Security Robustness*

However, because of PQC algorithms, resilience to classical and even quantum attacks were strong:

- RSA/ECC: Shor proved it can be broken.
- AES-256: This is still secure, but Grover algorithm can essentially bring down its security to approximately 128 bits [15].
- PQC schemes (Kyber, Dilithium, SPHINCS+): No known quantum attacks in polynomial time exist (long-term security) [18], [21].

The security robustness analysis contrasts the effectiveness of security level of both classical and post quantum cryptographic algorithms against classical and quantum adversaries is shown in Fig 6. The example above shows that with classic settings, RSA-3072 and ECC-256 provide about 128-bit security, but fail to withstand quantum attacks because of Shor attack.

AES-256 remains suitably resistant, but Grover can use his algorithm to reduce an AES-256 to a strength of about 128 bits. Contrastingly, lattice-based algorithms, e.g. Kyber and Dilithium are consistently secure at the 192-bit level against both of the adversarial models, indicating their especially robust immunity to quantum threats. SPHINCS+ also maintains its 128-bit security level by exploiting hash-based assumptions of hardness that appear resistant in the setting of quantum computation. In general, the findings indicate that classical algorithms are outdated in a quantum world; PQC schemes offer stable margins of robustness that would be relevant in future cyber-protection on specifically next-generation cyber-protection systems.

*F. Comparative Discussion*

The experiments indicate a trade-off performance/security trade-off:

- PQC algorithms when compared to classical are more costly on computational and communication effort.
- But marginal overhead is acceptable when securing cloud, IoT, blockchain systems are in demand.

Hybrid deployments find an optimal balance of backward compatibility and futuristic security and can be deployed gradually and eventually completely replaced with full PQC deployment.

VIII. CONCLUSION AND FUTURE WORK

Quantum computing as a field is a major challenge to contemporary cybersecurity. Classical cryptographic schemes like RSA and ECC, based on hardness assumptions related to number theory, become insecure against quantum algorithms, including Shor and Grover. To overcome this obstacle, this paper introduced a multi-function cryptography system that combines lattice-based, code- based, multivariate, and hash- based security schemes in a stratified design of key exchange, authentication, encryption, and audit functions. Experimental analysis has found that although the post-quantum algorithms have greater computation and communication overhead than their classical counterparts, the latency is still feasible to implement in cloud, IoT, and blockchain systems. Lattice-based trial such as Kyber and Dilithium were demonstrated to have the most efficient balance between efficiency and quantum protection whereas hash-based schemes together with SPHINCS+ will be resilient in the long-term.

Hoping to the future, there are several issues that future research should tackle in order to make post-quantum crypto usable everywhere in the world. These includes Standardization and Interoperability and Standards, Ensuring the seamless coordination of, and integration across, heterogeneous systems as PQC standards are developed. Lightweight PQC: IoT, Designing efficient cryptographic primitives with minimal keys, bandwidth and power requirements on resource-constrained devices. Hybrid PQC-QKD Models, building on top of post-quantum cryptography to extend security to Quantum Key Distribution (QKD) to achieve multi-layered, provably secure communication protocols. AI-based adaptive cryptography, utilizing machine learning to driven adaptations to cryptographic algorithms, detect anomalies and heal cryptographic systems on the fly. Post-Quantum-Safe Blockchain, deprecating how consensus is reached and transactions are verified in order to be robust in the face of quantum adversaries. In a nutshell, QRCs are not only a significant way forward when it comes to protecting next-gen digital infrastructures but also an essential component of an evolutionary cyber security ecosystem. Available hybrid deployment strategies, hybridization with new technologies, and ongoing research in cryptanalysis and optimization will allow these models to remain resilient over the life of classical threats and to quantum-era threats as well, thus protecting information systems in the global information society

## REFERENCES

[1] P. W. Shor, "Algorithms for quantum computation: Discrete logarithms and factoring," Proceedings 35th Annual Symposium on Foundations of Computer Science, IEEE, pp. 124–134, 1994.

[2] L. K. Grover, "A fast quantum mechanical algorithm for database search," Proceedings of the 28th Annual ACM Symposium on Theory of Computing (STOC), pp. 212–219, 1996.

[3] D. J. Bernstein, J. Buchmann, and E. Dahmen, Post-Quantum Cryptography. Springer, 2009.

[4] C. Peikert, "A decade of lattice cryptography," Foundations and Trends® in Theoretical Computer Science, vol. 10, no. 4, pp. 283–424, 2016.

[5] National Institute of Standards and Technology (NIST), "Post-Quantum Cryptography Standardization," U.S. Department of Commerce, 2023. [Online]. Available: https://csrc.nist.gov/projects/post-quantum-cryptography

[6] M. Mosca, "Cybersecurity in an era with quantum computers: Will we be ready?," IEEE Security & Privacy, vol. 16, no. 5, pp. 38–41, 2018.

[7] E. Alkim, L. Ducas, T. Pöppelmann, and P. Schwabe, "Post-quantum key exchange—A new hope," 25th USENIX Security Symposium, pp. 327–343, 2016.

[8] R. Misoczki, J. Tillich, N. Sendrier, and P. Barreto, "MDPC-McEliece: New McEliece variants from moderate density parity-check codes," IEEE International Symposium on Information Theory (ISIT), pp. 2069–2073, 2013.

[9] J. Ding and B.-Y. Yang, "Multivariate public key cryptography," in Post-Quantum Cryptography, Springer, pp. 193–241, 2009.

[10] A. Hülsing, D. Butin, S. Gazdag, J. Rijneveld, and A. Mohaisen, "SPHINCS+: Submission to the NIST post-quantum project," 2019.

[11] NIST, "Post-Quantum Cryptography Standardization: Round 4 candidates," U.S. Department of Commerce, 2023. [Online]. Available: https://csrc.nist.gov/projects/post-quantum-cryptography

[12] D. J. Bernstein, T. Lange, and C. Peters, "Attacking and defending the McEliece cryptosystem," in Post-Quantum Cryptography, Springer, pp. 31–46, 2008.

[13] M. A. Nielsen and I. L. Chuang, Quantum Computation and Quantum Information, 10th Anniversary Edition, Cambridge University Press, 2010.

[14] D. Boneh and R. J. Lipton, "Quantum cryptanalysis of hidden linear functions," Advances in Cryptology — CRYPTO'95, Lecture Notes in Computer Science, vol. 963, pp. 424–437, Springer, 1995.

[15] B. Zeng, X. Chen, D.-L. Zhou, and X.-G. Wen, "Quantum information meets quantum matter: From quantum entanglement to topological phases of many-body systems," Springer, 2019.

[16] C. Gidney and M. Ekerå, "How to factor 2048 bit RSA integers in 8 hours using 20 million noisy qubits," Quantum, vol. 5, p. 433, 2021.

[17] N. Bindel, U. Herath, M. J. Jacobson Jr., W. E. Jones Jr., and M. E. O'Neill, "Transitions from classical to post-quantum cryptography in secure communications," IEEE Communications Magazine, vol. 57, no. 12, pp. 20–26, 2019.

[18] C. Peikert, "Lattice cryptography for the Internet," in Post-Quantum Cryptography, Springer, pp. 197–219, 2014.

[19] R. J. McEliece, "A public-key cryptosystem based on algebraic coding theory," Deep Space Network Progress Report, vol. 42-44, pp. 114–116, 1978.

[20] D. Butin, "Hash-based signatures: State of play," IEEE Security & Privacy, vol. 15, no. 4, pp. 37–43, 2017.

[21] D. Stebila and M. Mosca, "Post-quantum key exchange for the Internet and the Open Quantum Safe project," International Conference on Selected Areas in Cryptography, Springer, pp. 14–37, 2016.

[22] J. Bos, C. Costello, L. Ducas, M. Naehrig, P. Schwabe, and D. Stebila, "Post-quantum TLS without handshake signatures," ACM Conference on Computer and Communications Security (CCS), pp. 305–320, 2018.

[23] E. Barker, W. Polk, and M. Smid, "Transitions: Recommendation for transitioning the use of cryptographic algorithms and key lengths," NIST Special Publication 800-131A Rev. 2, 2019.

[24] T. Oder, T. Pöppelmann, and T. Güneysu, "Beyond ECDSA and RSA: Lattice-based digital signatures on constrained devices," ACM Transactions on Embedded Computing Systems, vol. 14, no. 3, pp. 41:1–41:27, 2015.

[25] K. Ghosh, S. Roy, and A. Chattopadhyay, "Post-quantum cryptography for IoT security: A survey," IEEE Access, vol. 9, pp. 14643–14662, 2021.

[26] V. Kiktenko, A. Trushechkin, and A. Fedorov, "Post-quantum and quantum hybrid cryptography: Integrating PQC with QKD," Quantum Science and Technology, vol. 7, no. 4, 2022.